\documentclass[letterpaper, 10 pt, conference]{ieeeconf}  

\IEEEoverridecommandlockouts                              

\usepackage{amsfonts,amssymb,amsmath}

\usepackage{graphicx}
\usepackage{subcaption}

\title{\LARGE \bf
Generalized Multiple Correlation Coefficient as a Similarity Measurement between Trajectories
}

\author{ Julen Urain$^{1}$ \and Jan Peters$^{1,2}$
\thanks{
$^1$Intelligent Autonomous Systems, TU Darmstadt}
\thanks{$^2$MPI for Intelligent Systems, Tuebingen}
\thanks{The research leading to these results has received funding from the European Community’s Framework Programme SHAREWORK}
\thanks{\tt \small \{urain,peters\}@ias.tu-darmstadt.de}
}

\begin{document}

\maketitle
\thispagestyle{empty}
\pagestyle{empty}

\begin{abstract}

Similarity distance measure between two trajectories is an essential tool to understand patterns in motion, for example, in Human-Robot Interaction or Imitation Learning. 
The problem has been faced in many fields, from Signal Processing, Probabilistic Theory field, Topology field or Statistics field. Anyway, up to now, none of the trajectory similarity measurements metrics are invariant to all possible linear transformation of the trajectories~(rotation, scaling, reflection, shear mapping or squeeze mapping). Also not all of them are robust in front of noisy signals or fast enough for real-time trajectory classification. 
To overcome this limitation this paper proposes a similarity distance metric that will remain invariant in front of any possible linear transformation. Based on Pearson's Correlation Coefficient and the Coefficient of Determination, our similarity metric, the Generalized Multiple Correlation Coefficient~(GMCC) is presented like the natural extension of the Multiple Correlation Coefficient. The motivation of this paper is two-fold: First, to introduce a new correlation metric that presents the best properties to compute similarities between trajectories invariant to linear transformations and compare it with some state of the art similarity distances. Second, to present a natural way of integrating the similarity metric in an Imitation Learning scenario for clustering robot trajectories.
\end{abstract}

\section{INTRODUCTION}

In the industry of tomorrow it is expected the robots to collaborate with humans. The robots should not only imitate the humans~\cite{imit, imit2}, but also interact with us. The objective of Human-Robot Interaction~(HRI) is to understand intentions~\cite{predict1} and shape the interactions~\cite{general1} between one or more humans and one or more robots~\cite{HRI_survey}.

There is a lot of work done in human intention recognition and generalization. In~\cite{amor1}, Amor et al. developed a Human-robot solution based on PPCA. Their algorithm computes a low-dimensional projection to encapsule human-robot interactions in the training and then in testing, observing human motion, the robot was able to infer his movement. They found out that for certain motions, even if the robot was taught how to interact when human was moving hands up, the robot was able to generalize even when the human hands where moving down. In~\cite{maeda1}, Interaction Primitives~(IP)~\cite{amor2} were used to learn human robot interactions. Probabilistic Motion Primitives~(PROMP)~\cite{parachos1} were used to model human motions for each skill and then the robot was able to guess the skill human was doing through Bayes and then generalize on the skill applying conditioning over PROMP. In~\cite{gomez1}, PROMP's were also used to adapt robot movements to a human-robot table tennis match. In~\cite{t_shi}, HMM were used to recognize human gestures. The recognition was invariant to the starting position of the gestures. The sequences of human motion were segmented into atomic components and clustered by HMM.

Anyway, up to now, the generalization on human gestures recognition has been applied on a few specific tasks and very limited to small variations between the learned human motion and the observed one. The actual algorithms tends to fail recognizing gestures if these are similar shape to the ones already in our gestures library but with different rotations.

This limitation can be overcome developing a good similarity measurement that can extract the relevant features from the learned human motions and then be able to recognize rotated or scaled new observations. In an ideal case of HRI, it should be enough teaching the robot how to interact with the human for some particular cases and then, during the testing, when the human is doing a similar shape motion, but maybe rotated, the robot could find the similarities between the previously learned motions and the new ones.

\begin{figure}[t]
  \includegraphics[width=\linewidth]{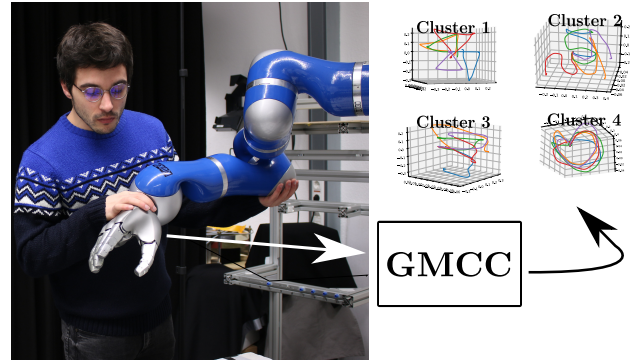}
  \caption{GMCC applied for clustering taught trajectories in Imitation Learning }
  \label{iros11}
\end{figure}

\subsection{Problem Statement}\label{prob_state}
The motion similarity has been studied in different fields. For example, In Probabilistic Theory field, the motion has been represented as a probability distributions. This let the similarity distance to be robust with noisy motions. In Signal Processing field the motion are represented as time series. Similarity metrics deals with time morphed time series. None of the commented similarity metrics consider spatial transformations among trajectories and so, these similarity metrics give poor performance measuring the  similarity distance among rotated or scaled trajectories.

Statistics field provides the best similarity metrics to compare spatially transformed trajectories. In Statistics, two trajectories similarity depends on how correlated are between each other. A high correlation between trajectories will mean a high similarity. For the case of unidimensional trajectories $x_t \in {\rm I\!R}$ and $y_t \in {\rm I\!R}$ where $t$ is the time samples, Pearson's correlation coefficient is measured as follows 
\begin{equation}
    \rho(x_t,y_t)=\frac{\textrm{cov}(x_t,y_t)}{\sigma_x\sigma_y} .
\end{equation}
The correlation coefficient will be 1 if the trajectories are totally positive correlated, 0 if there is no any linear correlation and -1 if there is a total negative linear correlation.
When it comes multidimensional variables, several correlation coefficients exist. The first multidimensional coefficient was proposed by Harold Hotelling in 1936~\cite{hotelling1}. The Canonical Correlation Coefficient~(CAC), measures the correlation between a linear combination of the first variable $\boldsymbol{X_t} \in {\rm I\!R}^{n}$ and a linear combination of second variable $\boldsymbol{Y_t} \in {\rm I\!R}^{n}$

\begin{align}
\label{eqn:eqlabel}
\begin{split}
   (\boldsymbol{a^*},\boldsymbol{b^*}) = \arg \max_{\boldsymbol{a},\boldsymbol{b}}\rho(\boldsymbol{a^\intercal X_t},\boldsymbol{b^\intercal Y_t}) ,
\\
 \textrm{CAC}(\boldsymbol{X_t},\boldsymbol{Y_t}) = \rho(\boldsymbol{a^{*\intercal} X_t},\boldsymbol{b^{*\intercal} Y_t}) .
\end{split}
\end{align}
Nowadays, the most famous multidimensional correlation coefficients for considering spatial transformations are two: RV~Coefficient and Distance Correlation~(dCor). RV Coefficient~\cite{escoufier1} is computed as follows
\begin{equation}
    RV(\boldsymbol{X_t},\boldsymbol{Y_t}) = \frac{tr(\boldsymbol{X_tX_t^\intercal Y_tY_t^\intercal})}{\sqrt{tr(\boldsymbol{X_tX_t^\intercal})^{2}tr(\boldsymbol{Y_tY_t^\intercal})^{2}}}.
\end{equation}
As it can be seen RV coefficient is computed with $\boldsymbol{X_tX_t^\intercal}$ and  $\boldsymbol{Y_tY_t^\intercal}$. If we consider $\boldsymbol{X_2} = \boldsymbol{XQ}$ where $\boldsymbol{Q}$ is any orthogonal matrix, then, $\boldsymbol{X_2X_2^\intercal} = \boldsymbol{XQQ^\intercal X^\intercal} = \boldsymbol{XX^\intercal} $. From here, it is concluded that RV coefficient will be invariant to any orthogonal transformation, so invariant to rotations.

On the other hand, Distance Correlation~\cite{dcov} is not only invariant to linear transformations but also to some non linear transformations also. Having two multidimensional vectors $\boldsymbol{X_t} \in {\rm I\!R}^{n}$ and $\boldsymbol{Y_t} \in {\rm I\!R}^{n}$; first, the euclidean distance is computed between different time samples
$$a_{k,j} = || \boldsymbol{X_j} - \boldsymbol{X_k} ||, \quad k,j = 0,\dots, T$$  
$$b_{k,j} = || \boldsymbol{Y_j} - \boldsymbol{Y_k} ||, \quad k,j = 0,\dots, T$$
where $||\cdot||$ is the euclidean distance. Once computed, $a_{k,j}$ and $a_{k,j}$ are double normalized
$$A_{j,k} = a_{k,j} - \overline{a}_{\cdot,k} - \overline{a}_{j,\cdot} + \overline{a}_{\cdot,\cdot}$$
$$B_{j,k} = b_{k,j} - \overline{b}_{\cdot,k} - \overline{b}_{j,\cdot} + \overline{b}_{\cdot,\cdot}$$
and the obtained values used for computing the distance covariance~(dCov) and the distance correlation~(dCor)
\begin{align}
\label{eqn:eqlabel2}
\begin{split}
\textrm{dCov}^{2}(\boldsymbol{X},\boldsymbol{Y}) = \frac{1}{n^2}\sum_{k=1}^n\sum_{j=1}^nA_{j,k}B_{j,k},
\\
\textrm{dCor}(\boldsymbol{X},\boldsymbol{Y}) = \frac{\textrm{dCov}(\boldsymbol{X},\boldsymbol{Y})}{\sqrt{\textrm{dVar}(\boldsymbol{X})\textrm{dVar}(\boldsymbol{Y})}}
\end{split}
\end{align}
where, $\textrm{dVar}^2(\boldsymbol{X}) =\textrm{dCov}^2(\boldsymbol{X},\boldsymbol{X}) $. dCor will only consider the sign of the distances computed between different time samples. If the normalized distances from $k$ to $j$ have the same sign in both $A_{kj}$ and $B_{kj}$, the correlation will increase and if the signs are different then decrease. So, dCor does not care if the relation between the variables is linear or nonlinear.

In this paper, we propose a new correlation measurement for multidimensional variables. As far as we know, there is no any trajectory similarity measurement that remains invariant to any linear transformations and only linear transformations. Our algorithm is computed as an extension of the Multiple Correlation Coefficient for multidimensional variables and it will be demonstrated that the Multiple Correlation Coefficient can be considered as a particular case of the the Generalized Multiple Correlation Coefficient(GMCC) for unidimensional variables.

The rest of the paper is organized as follows: In Section~\ref{RW}, similarity measures from other fields are presented. In Section~\ref{GMCC}, first the mathematical basis of our algorithm is briefly explained, to continue with the presentation of the novel similarity metric, GMCC. Section~\ref{exp} is dedicated to the experiments developed on the robot. First, GMCC's characteristics are studied and compared to RV~Coefficient and dCor. Later, the GMCC is applied in a clustering task of robot trajectories~Fig.~\ref{iros11}. Finally, the paper concludes in Section~\ref{conclusions} discussing ideas for future work. 

\subsection{Related Work}\label{RW}

The similarity measurements presented in Section~\ref{prob_state} are the best similarity measurements when it comes to measure spatial invariant similarities. Nevertheless, there are also relevant similarity measurement in other fields: in \ref{SIGNAL}) similarity distances from Signal Processing field are presented, in \ref{TOPOLOGY}) similarity distances from Topology field and in \ref{PROBAB}) similarity distances from Probabilistic Theory field.

 \begin{enumerate}
   \item In \textbf{Time Series} field, the similarity has been faced from time shifting perspective. Dynamic Time Warping~(DTW)~\cite{dtw1} is a well-known algorithm. Considering two similar trajectories, DTW finds the best matching between them applying time shifting, anyway it doesn't give any metric about how similar two time series are. A similarity measure considering the time shifting is the Edit Distance With Real Penalty~(EDR)~\cite{EDR}. On it, Chen~et~al. propose a similarity measurement based on string edit distance. The similarity distance between two trajectories will be related with the number of elements we should add or delete to make them equal. None of these similarity metrics are invariant to spatial transformations. \label{SIGNAL}
   \item \label{TOPOLOGY}
On \textbf{Topology} field, the similarity between trajectories is studied without considering the time information and so, only spatial information is used. The two most well-known similarity measurements are Frechet distance~\cite{Frechet} and Hausdorff distance~\cite{hausdorff}. Frechet distance is presented as the bottleneck maximun distance between two trajectories. From a different approach, Hausdorff distance is the greatest of all the distances from a point in one set to the closest point in the other set. These distances doesn't consider spatial transformations of the trajectories.
\item \label{PROBAB}
On \textbf{Probabilistic Theory} field, Maximum Likelihood is the most used algorithm for measuring the similarity between two trajectories. In~\cite{predict1}, Maximum Likelihood was used to compute the similarity between learned gestures and the observations. The benefits of using probabilistic approaches are several. In contrast with similarity metrics from other fields, Maximum Likelihood can not only give the probability of how similar two trajectories are, but also, predict the trajectory from partial observations of the trajectory. So, the algorithms can be used for recognizing and predicting the human motions. The drawback is that no spatial transformations are considered between learned motions and the observed ones.
 \end{enumerate}

\section{Generalized  Multiple  Correlation  Coefficient}\label{GMCC}

We introduce a new correlation coefficient. This coefficient can be used as a similarity measure between multidimensional time series. Having two multidimensional trajectories $\boldsymbol{X_t} \in {\rm I\!R}^{n}$ and $\boldsymbol{Y_t} \in {\rm I\!R}^{n}$ where $t$ is the time samples and n the dimensions, the Generalized Multiple Correlation Coefficient~(GMCC)  is a measure of how well can be mapped a linear transformation between the trajectories $\boldsymbol{X_t}$ and $\boldsymbol{Y_t}$. As lot of correlation measurements, the GMCC takes values between 0 and 1. A coefficient of 1 means that there is a complete correlation between $\boldsymbol{X_t}$ and $\boldsymbol{Y_t}$, and a value of 0 means that there is no any possible linear relation between the vectors $\boldsymbol{X_t}$ and $\boldsymbol{Y_t}$.

The invariability on linear transformation is the biggest contributions of this new similarity metric. The coefficient will remain invariant in front of any possible linear transformation and only linear transformations, so the correlation coefficient will decrease sharply when no linear correlations appear. This contribution is particularly interesting in comparison with other correlation coefficients such as RV coefficient, that remains invariant with only with orthogonal transformation matrices and dCor that remains invariant with not only linear but also some nonlinear transformations.

As additional contributions, and similarly to other correlation coefficients, GMCC is robust when noisy trajectories are compared and the coefficient is obtained instantly in comparison with other Signal Processing field similarity metrics such as EDR~\cite{EDR} or TWED~\cite{TWED} which computation time is of $\mathcal{O}(t^2)$ where t is the time samples of the compared trajectories $\boldsymbol{X_t}$ and $\boldsymbol{Y_t}$.

GMCC is an extension of the Multiple Correlation Coefficient. While the Multiple Correlation Coefficient computes the correlation between an univariate dependant variable $y$ and multiple independent variables $\boldsymbol{X}$, the GMCC extends it to multivariate dependant variables $\boldsymbol{Y}$. Moreover, in Section~\ref{GMCC2} it will be demonstrated that Multiple Correlation Coefficient can be computed as a particular case of the GMCC.

The Section~\ref{GMCC} continues as follows: In Section~\ref{GMCC1} a brief introduction to the mathematical basis for developing our coefficient is presented. Once the Coefficient of Determination and the Multiple Correlation Coefficients are introduced, in Section~\ref{GMCC2} our novel coefficient, GMCC is presented and computed.

\subsection{Coefficient of Determination}\label{GMCC1}
Considering an univariate dependant variable $y_t \in {\rm I\!R}$ and multiple independent variables $\boldsymbol{X_t} \in {\rm I\!R}^{n} $, the coefficient of determination, also know as $R^2$ is the fraction of the variance in the dependent variable~$y$ that is predictable from the independent variables~$\boldsymbol{X}$~\cite{mc}

\begin{equation}
    R^2 = \frac{\sum_{t=0}^T (\hat{y}_t -\overline{y})^2}{\sum_{t=0}^T(y-\overline{y})^2}
    \label{r2}
\end{equation}
where $\hat{y}_t = \boldsymbol{A^{\intercal}X} = a_1x_{1t} + a_2x_{2t} + \dots +  a_nx_{nt} $ with $\boldsymbol{A} \in \mathbb{R}^n$ is the prediction model. $y_t$ can be expressed in terms of predicted values $\hat{y}_t$ and residual errors $\epsilon_t$
$$y_t = \hat{y}_t + \epsilon_t $$
then, the parameters $\boldsymbol{A}$ that minimize the squared sum of the residual error $\epsilon_t$ can be obtained through an optimization function
\begin{equation}
    \boldsymbol{A^*} = \arg\min_{\boldsymbol{A}}(\sum_{t=1}^T \epsilon_t^2)
    \label{AAA}
\end{equation}
The solution of the optimization function from Eq.~\ref{AAA}, $\boldsymbol{A^*}$, is computed by Least-Squared regression and so, the optimal prediction model is $\hat{y}_t^* = \boldsymbol{A^{*\intercal}X_t}$. In the particular case of having this optimal model, the Coefficient of Determination can be expressed in terms of the Pearson's Correlation coefficient~\cite{mc}
\begin{equation}
    R^2 = \boldsymbol{c^\intercal R_{xx}^{-1}c^\intercal}
\label{MCC}
\end{equation}
where $\boldsymbol{c} =(\rho_{yx_1},\dots, \rho_{x_n}) $ is a vector with the correlation coefficient between each individual independent variables~$x_i$ and the dependent variable~y. $\boldsymbol{R_{xx}}$ is a correlation matrix measuring the correlation between the independent variables
$$\boldsymbol{R_{xx}} = \begin{bmatrix}
    \rho_{11} & \rho_{12} & \rho_{13} & \dots  & \rho_{1n} \\
    \rho_{21} & \rho_{22} & \rho_{23} & \dots  & \rho_{2n} \\
    \vdots & \vdots & \vdots & \ddots & \vdots \\
    \rho_{n1} & \rho_{n2} & \rho_{n3} & \dots  & \rho_{nn}
    \end{bmatrix}.$$
It is important to remark that in the case of having the independent variables completely uncorrelated between each other,$\boldsymbol{R_{xx}}$ becomes into the identity matrix

\begin{equation}
    \boldsymbol{R_{xx}} = \boldsymbol{\mathbb{I}_{n\textrm{x}n}}
    \label{identity}
\end{equation}
and so, the coefficient of determination can be expressed as a sum of squared Pearson's Correlation coefficients. Mixing Eq.~\ref{MCC} and Eq.~\ref{identity}
$$R^2 = \sum_{i=1}^n \rho_{yx_i}^2.$$
Multiple Correlation Coefficient is the square root of Eq.~\ref{MCC} and it represents how well we can predict the parameters of the dependant variable $y$ with the multiple independent variables $\boldsymbol{X}$ using the optimal linear model.
\begin{equation}
    R = \sqrt{\boldsymbol{c^\intercal R_{xx}^{-1}c^\intercal}}
\end{equation}

As remarked in~\cite{mc}, Multiple Correlation Coefficient will increase when the number of the independent variables increase and so it will give a high correlation coefficient between any two variables. From regression perspective is logical. As we are applying Least-Square regression to fit $\boldsymbol{X}$ with $y$, when the independent variable increase, the noise can be used to fit in y and so, we could have over-fitting problems. Anyway, it is expected the noise not to affect a lot in low-dimensional cases such as three dimensional euclidean motions.

\subsection{Generalized Multiple Correlation Coefficient}\label{GMCC2}

We propose a new correlation coefficient for the case of multiple dependant variables $\boldsymbol{Y_t} \in {\rm I\!R}^{n}$ and multiple independent variables $\boldsymbol{X_t} \in {\rm I\!R}^{n}$ where $t$ is the time sample and $n$ is the dimension. The given name is Generalized Multiple Correlation Coefficient~(GMCC), as it can be introduced as the extension of the Multiple Correlation Coefficient for multivariate dependent variables. The most relevant property from GMCC in comparison with the state of the art multivariate correlation coefficients such as RV and dCor is that it will remain invariant to any possible linear transformations and only linear transformations. For any nonsingular matrix~$\boldsymbol{H}$
$$\textrm{GMCC}(\boldsymbol{X},\boldsymbol{HX})=1.$$ 
For Human-Robot Interaction framework or Imitation Learning framework, a similarity measurement that remains invariant between linearly transformed trajectories is a useful tool. This is the case of different scale, rotation, mirror, squeeze mapped or shear mapped trajectories. This similarity measurement can be applied from classication of human gestures in HRI environment to clustering of teached skills in Imitation Learning as we will show in Section~\ref{exp}.

GMCC lays between the  RV coefficient that remains invariant in front of rotations and scaling and the distance correlation~(dCor) that remains invariant for not only linear transformations but also some nonlinear transformations. In order to obtain a scalar correlation coefficient between $\boldsymbol{Y}$ and $\boldsymbol{X}$ multidimensional trajectories, we propose a modified Coefficient of Determination. Based on Eq.~\ref{r2}
\begin{equation}
        R^2 = \frac{\sum_{t=0}^T ||\boldsymbol{\hat{Y}_t} -\overline{\boldsymbol{Y}}||^2}{\sum_{t=0}^T||\boldsymbol{Y_t}-\overline{\boldsymbol{Y}}||^2}
    \label{r22}
\end{equation}
where $||\cdot||$ is the euclidean norm and $\boldsymbol{\hat{Y}_t} = \boldsymbol{HX}$ where $\boldsymbol{H}$ is any nonsingular matrix. The $R^2$ coefficient from Eq.~\ref{r22} is presented as the generalization of the Coefficient of Determination. Moreover, $R^2$ coefficient from Eq.~\ref{r2}, can be computed as a particular case of Eq.~\ref{r22} for unidimensional variable $y$.

If Eq.~\ref{r22} is extended
$$R^2=\frac{\sum_{t=0}^T\sqrt{(\hat{y}_{1t} - \overline{y}_1)^2 + \dots + (\hat{y}_{nt} - \overline{y}_n)^2}^2}{\sum_{t=0}^T\sqrt{(y_{1t} - \overline{y}_1)^2 + \dots + (y_{nt} - \overline{y}_n)^2}^2}$$
and now, if square root is eliminated by the power in both the numerator and denominator and the numerator split
$$R^2=\frac{\sum_{t=0}^T(\hat{y}_{1t} - \overline{y}_1)^2}{\sigma_Y^2} + \dots + \frac{\sum_{t=0}^T(\hat{y}_{nt} - \overline{y}_n)^2}{\sigma_Y^2}$$
where $\sigma_Y^2 = \sum_{t=0}^T(y_{1t} - \overline{y}_1)^2 + \dots + \sum_{t=0}^T(y_{nt} - \overline{y}_n)^2 = \sigma_{y_1}^2 + \dots +\sigma_{y_n}^2 $. If now, each fraction is multiplied by their respective variance of the dependant variable $\sigma_{y_i}^2$, the obtained equation is
$$R^2 = \frac{\sum_{t=0}^T(\hat{y}_{1t} - \overline{y}_1)^2}{\sigma_{y_1}^2}\frac{\sigma_{y_1}^2}{\sigma_Y^2} + \dots + \frac{\sum_{t=0}^T(\hat{y}_{nt} - \overline{y}_n)^2}{\sigma_{y_n}^2}\frac{\sigma_{y_n}^2}{\sigma_Y^2}.$$
For the particular case of the predicted linear models $\hat{y}_{it}$ been optimal, based on Eq.~\ref{MCC}, it can be rewritten as a weighted sum of the square of the Multiple Correlation Coefficient of each variable of $\boldsymbol{Y}$
\begin{equation}
    R^2 = \sum_{i=1}^n R_{y_i}^2\frac{\sigma_{y_i}^2}{\sigma_Y^2}
    \label{optimalr22}
\end{equation}
General Multiple Correlation Coefficient~(GMCC) is the square root of the general case of Coefficient of Determination from Eq.~\ref{r22} under the particular case of optimal linear prediction model computed in Eq.~\ref{optimalr22}
\begin{equation}
    R = \sqrt{\sum_{i=1}^n R_{y_i}^2\frac{\sigma_{y_i}^2}{\sigma_Y^2}}.
\end{equation}
It can be observe, that for the particular case of $\boldsymbol{Y}$ being unidimensional, The General Multiple Correlation Coefficient becomes into the Multiple Correlation Coefficient.

The Generalized Multiple Correlation Coefficient represents how well a set of independent multidimensional variables can predict a set of dependent multidimensional variables by a linear model. GMCC will take values between 0 and 1, where 1 means a complete prediction of $\boldsymbol{Y}_t$ with $\boldsymbol{X}_t$. One of the drawbacks of GMCC in comparison with RV coefficient and dCor coefficient is that it is not symmetric so, $R(\boldsymbol{X},\boldsymbol{Y}) \neq R(\boldsymbol{Y},\boldsymbol{X})$. In order to solve it, we can compute the symmetric GMCC
\begin{figure*}[t]
\centering
\begin{subfigure}{.5\textwidth}
  \includegraphics[width=.8\linewidth]{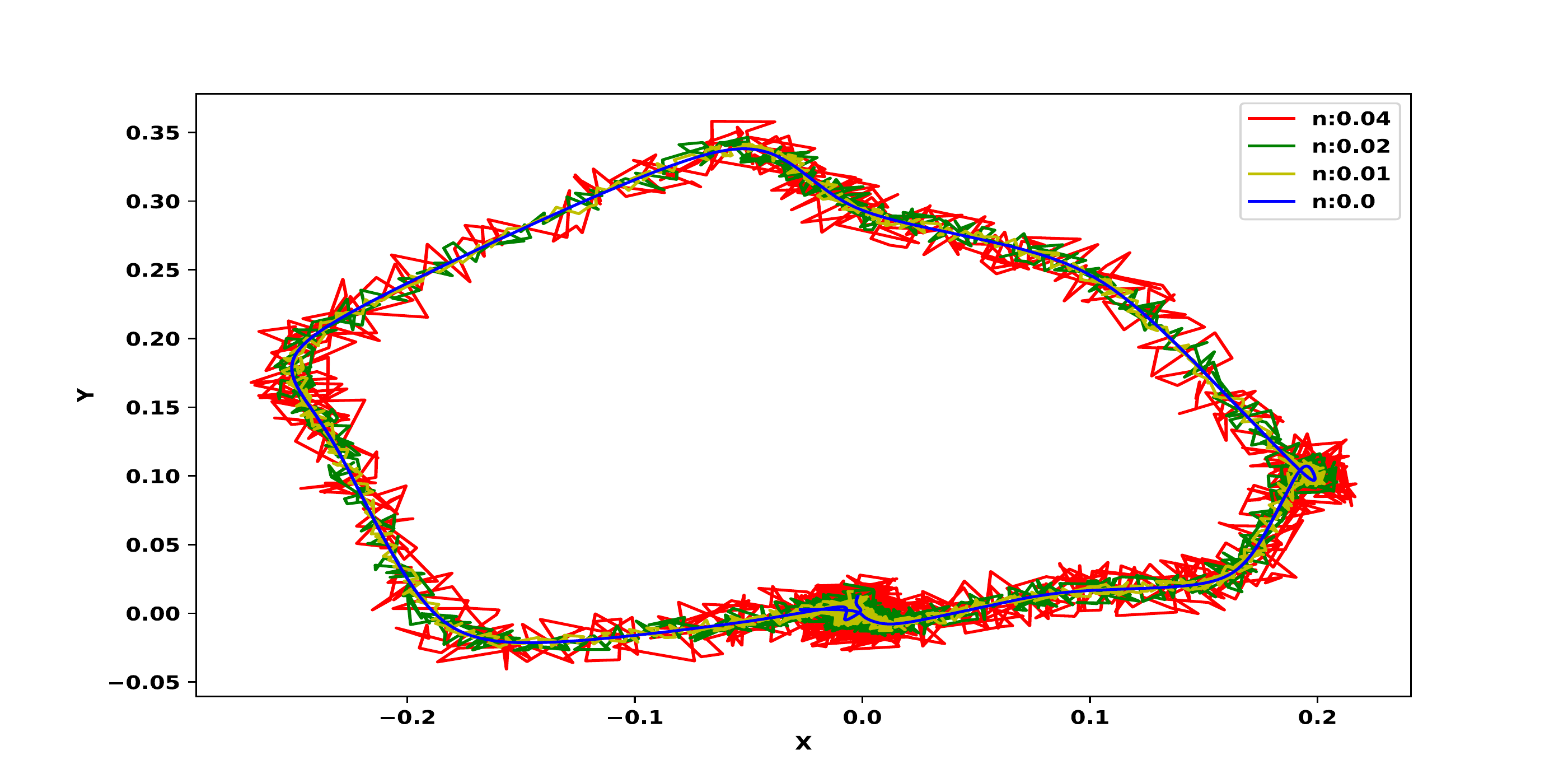}
  \caption{Circular trajectory with artificial gaussian noise}
  \label{fig:cn}
\end{subfigure}%
\begin{subfigure}{.5\textwidth}
  \includegraphics[width=.8\linewidth]{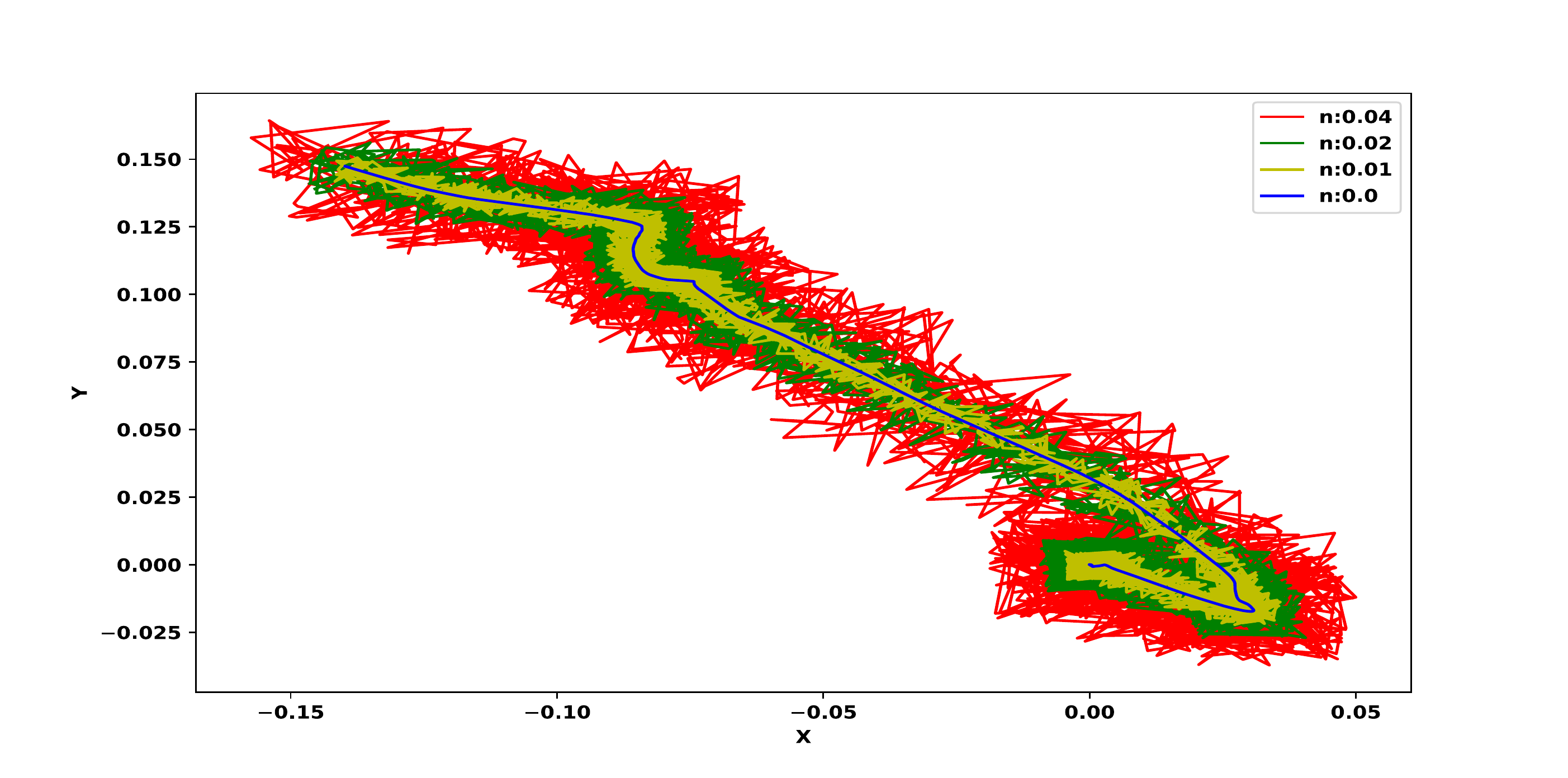}
  \caption{Linear trajectory with artificial gaussian noise}
  \label{fig:ln}
\end{subfigure}
\begin{subfigure}{.5\textwidth}
  \includegraphics[width=.8\linewidth]{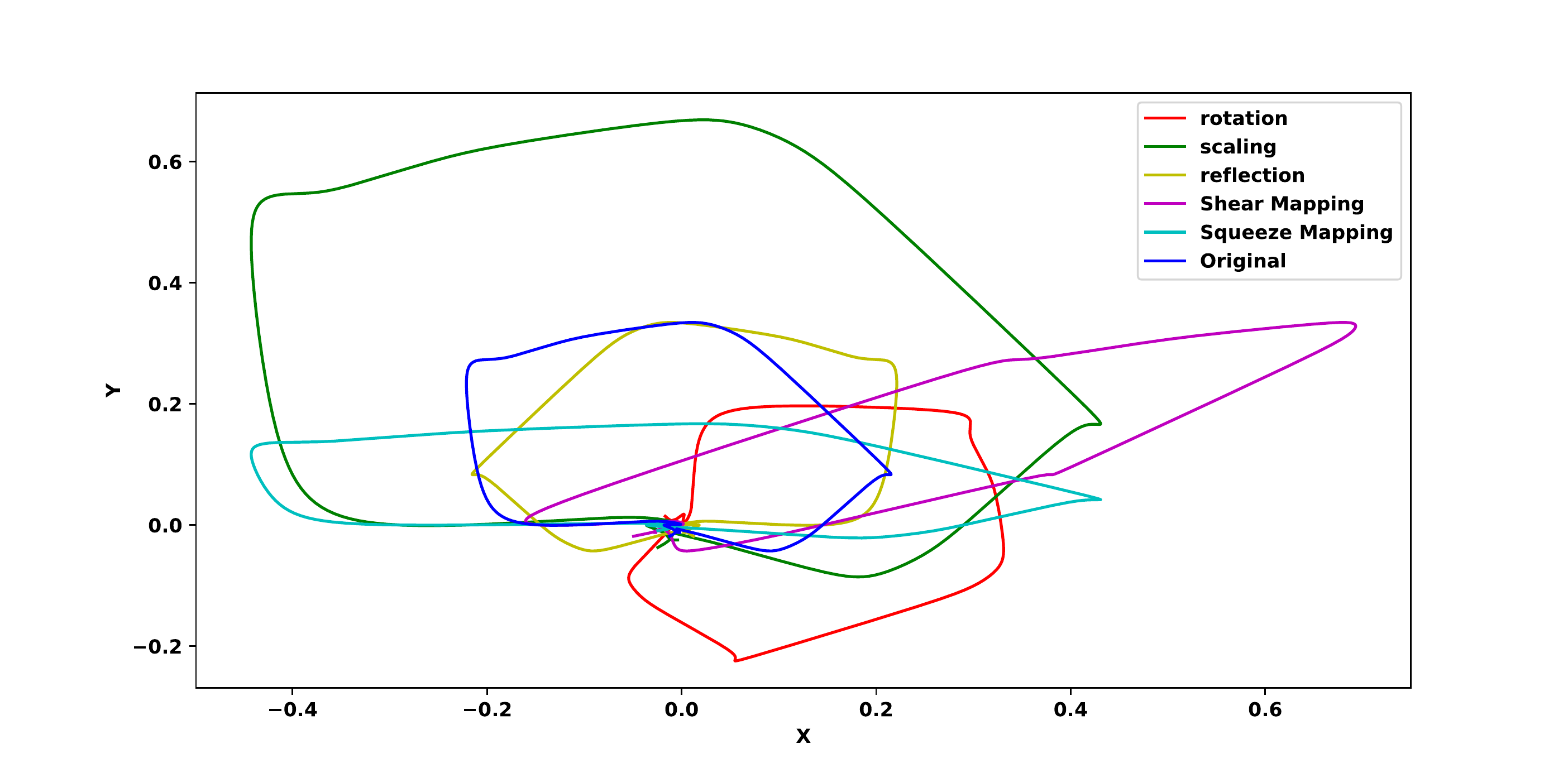}
  \caption{Circular trajectories artificially linearly transformed}
  \label{fig:cr}
\end{subfigure}%
\begin{subfigure}{.5\textwidth}
  \includegraphics[width=.8\linewidth]{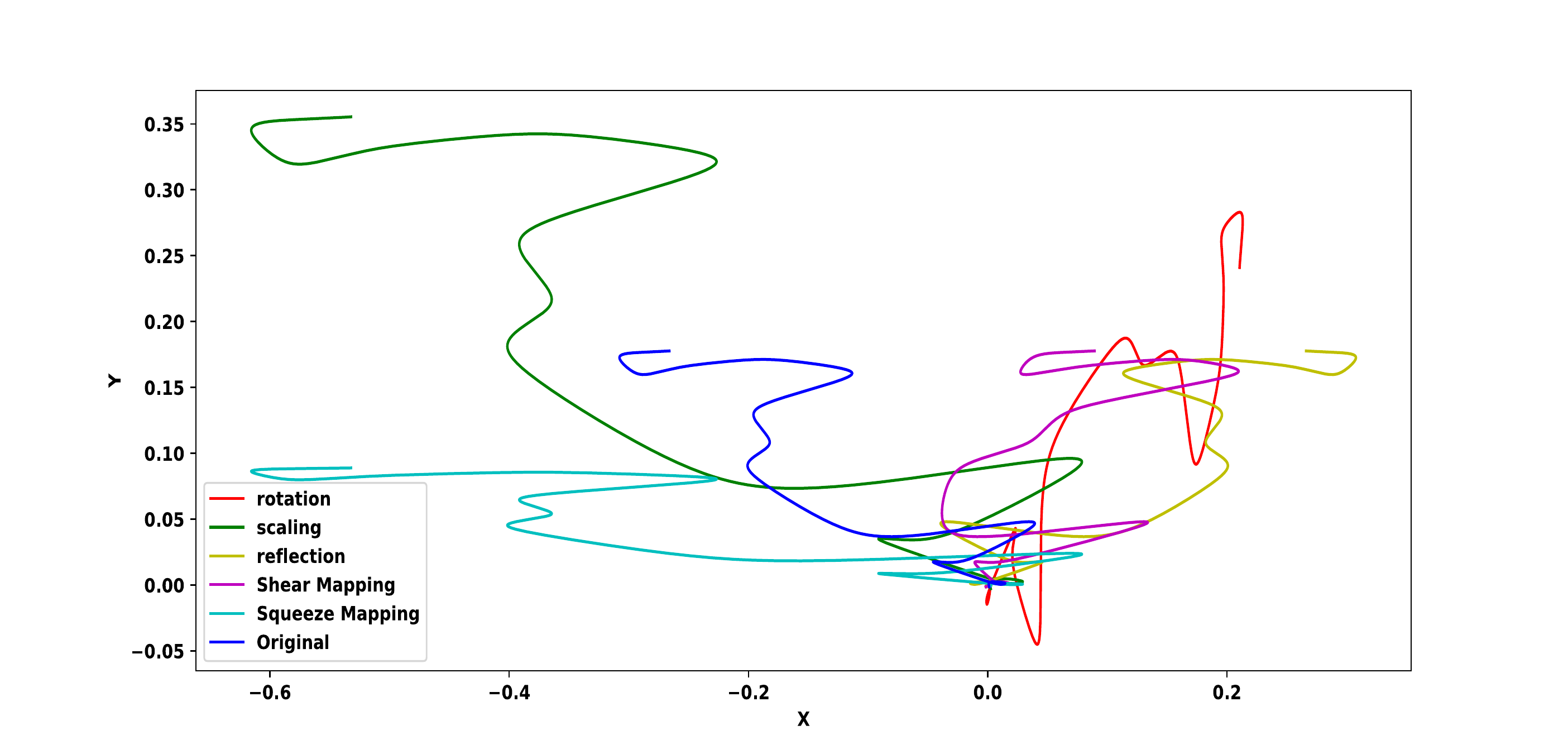}
  \caption{Linear trajectories artificially linearly transformed}
  \label{fig:lr}
\end{subfigure}
\caption{Artificially generated trajectories for testing the experiments in Section \ref{characteristics}}
\end{figure*}

\begin{table}[b]
\centering
\begin{tabular}{ |c|c|c|c| } 
\hline
Traj. type & dCor & RV & GMCC \\
\hline
line 0.01 & 0.0006 & 0.0 & 0.001 \\ 
line 0.02 & 0.02 & 0.0 &0.005  \\ 
line 0.04 & 0.01 & 0.0 & 0.02 \\ 
circle 0.01 & 0.0 & 0.0 & 0.001\\ 
circle 0.02 & 0.0 & 0.0 & 0.002\\ 
circle 0.04 & 0.0035 & 0.0 & 0.0039 \\ 
\hline
\end{tabular}
\caption{Similarity distance measurements for line and circular trajectories with different noises for distance correlation(dCor), RV coefficient and GMCC}
\label{noise}
\end{table}

\begin{equation}
    R_{sym}(X,Y) = \frac{R(X,Y) + R(Y,X)}{2}
    \label{sym}
\end{equation}
This symmetric correlation coefficient from Eq.~\ref{sym} is used to compute a similarity distance measure
\begin{equation}
    d_{X,Y} = 1-R_{sym}(X,Y).
    \label{distance}
\end{equation}
A similarity distance of 1 means that $\boldsymbol{X}$ and $\boldsymbol{Y}$ are completely uncorrelated and the closer the distance is to 0, the higher the linear correlation between $\boldsymbol{X}$ and $\boldsymbol{Y}$.

In the experimental evaluation, this distance measure is going to be compared with the two most relevant multivariate correlation coefficients metrics RV and dCor.

\section{EXPERIMENTAL EVALUATION}\label{exp}

For the experimental evaluation of GMCC, the experiments have been split in two sections. In Section \ref{characteristics}, the GMCC characteristics will be studied in some predefined trajectories. This Section will be useful to facilitate the reader the understanding of the capacities and limitations of the novel metric.

In Section \ref{toy_traj}, GMCC is applied for a trajectory clustering problem and the obtained solution compared with RV and dCor solutions. Six different type of trajectories have been recorded with a KUKA LWR arm. The recorded trajectories have been clustered through a hierarchical clustering method.

\subsection{GMCC characteristics}\label{characteristics}

For this first evaluation, two simple 2D signals are going to be used: a linear and a circular trajectory. To observe the robustness of GMCC in front of noisy signals, different variance gaussian noises were added in both the linear and circular trajectories. The obtained noisy trajectories are presented in Fig.~\ref{fig:cn} and Fig.~\ref{fig:ln}. The similarity distances between the original trajectories and the noisy signals were computed for RV, dCor and GMCC and the obtained values are presented in Table \ref{noise}.

\begin{table}[b]
\centering
\begin{tabular}{ |c|c|c|c| } 
\hline
Transf. type & dCor & RV & GMCC \\
\hline
rotation    & 0.0 & 0.01 & 0.0 \\ 
scale       & 0.0 & 0.0 & 0.0  \\ 
reflection  & 0.0 & 0.009 & 0.0 \\ 
shear       & 0.12 & 0.18 & 0.0\\ 
squeeze     & 0.11 & 0.14 & 0.0\\ 
\hline
\end{tabular}
\caption{Similarity distance measurements for circular trajectories with different transformations for distance correlation(dCor), RV coefficient and GMCC}
\label{transf_circle}
\end{table}

From Table \ref{noise}, it can be observed that in terms of noise the three similarity distances are pretty robust. GMCC is giving the worst results in comparison with dCor and RV, but from a general perspective, three of them remains robust in front of noisy signals as the obtained values are close to 0.

In a second experiment the selected two trajectories, a line and a circle, were transformed by rotation, scaling, reflection, shear mapping and squeezing. The obtained trajectories are presented in Fig.~\ref{fig:cr} and Fig.~\ref{fig:lr}. As the previous case, the similarity distances were computed between the original trajectory and the transformed ones for dCor, RV and GMCC and obtained results organized in Tables \ref{transf_circle} and \ref{transf_line}.
\begin{figure*}[t]
\centering
\begin{subfigure}{.33\textwidth}
  \includegraphics[width=\linewidth]{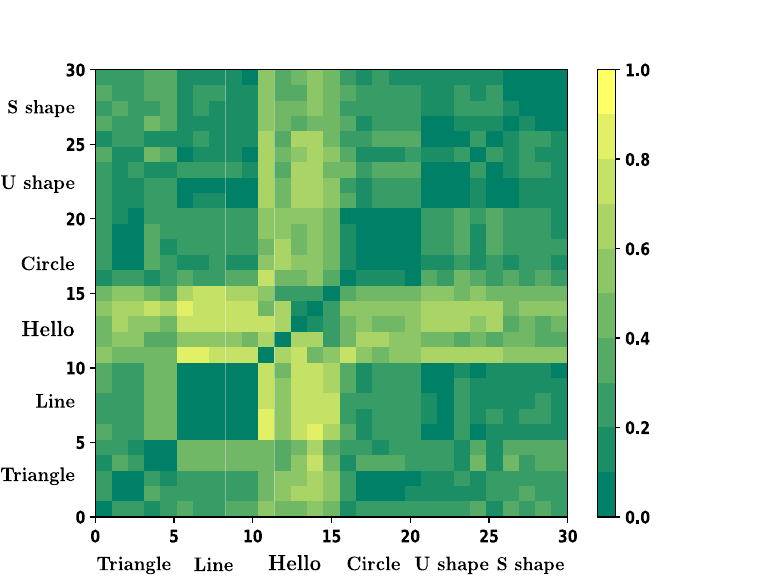}
  \caption{RV Coefficient distance matrix}
  \label{fig:simmrv}
\end{subfigure}%
\begin{subfigure}{.33\textwidth}
  \includegraphics[width=\linewidth]{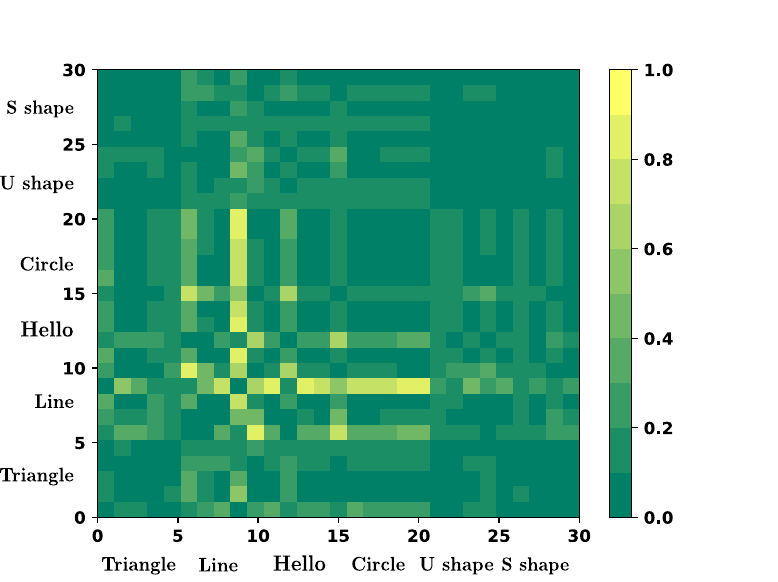}
  \caption{dCor distance matrix}
  \label{fig:simmdcor}
\end{subfigure}
\begin{subfigure}{.33\textwidth}
  \includegraphics[width=\linewidth]{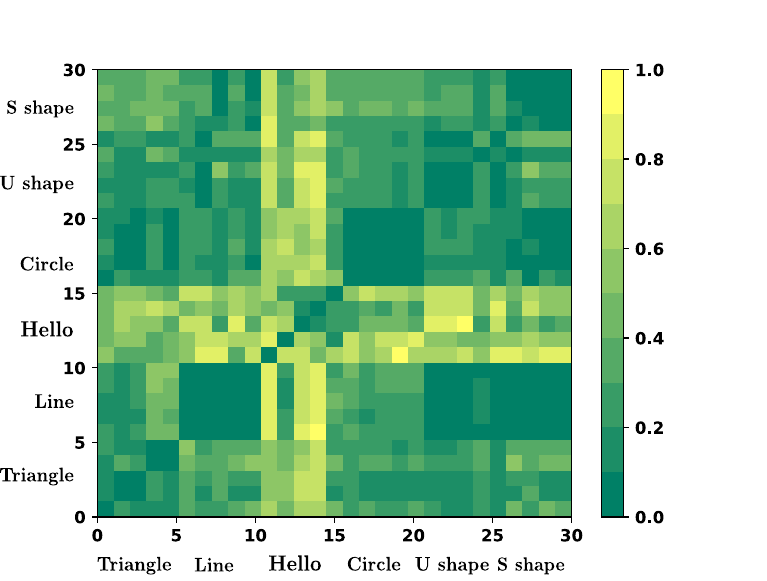}
  \caption{GMCC distance matrix}
  \label{fig:simmgmcc}
\end{subfigure}%
\caption{Distance Matrices computed for RV, dCor and GMCC}
\label{fig:siminar_mat}
\end{figure*}

\begin{table}[b]
\centering
\begin{tabular}{ |c|c|c|c| } 
\hline
Transf. type & dCor & RV & GMCC \\
\hline
rotation    & 0.0 & 0.82 & 0.0 \\ 
scale       & 0.0 & 0.0 & 0.0  \\ 
reflection  & 0.0 & 0.819 & 0.0 \\ 
shear       & 0.007 & 0.88 & 0.0\\ 
squeeze     & 0.0029 & 0.16 & 0.0\\ 
\hline
\end{tabular}
\caption{Similarity distance measurements for linear trajectories with different transformations for dCor, RV and GMCC}
\label{transf_line}
\end{table}

These tables are representative of the real power of GMCC. While the RV coefficient is measuring big distances between the original line and the one transformed by shear mapping, rotation or squeezing, the GMCC remains robust in null distance. dCor's distance is also pretty close to 0, but in shear mapping and squeezing gives a little bit worse results. GMCC is the coefficient that works better for similarity invariant to linear transformations and will remain always in lowest distance for any linear transformations.

GMCC is limited to linear transformations. GMCC distance starts increasing when no linear mapping can be set between the compared trajectories. Meanwhile, dCor is capable of finding also nonlinear mappings between trajectories. The distance measured by dCor will be smaller than  the distance with GMCC for nonlinear correlations. In order to prove this, the similarity distance has been measured between the original line and the original circle and the obtained distances presented in Table \ref{linvscir}.

\begin{table}[b]
\centering
\begin{tabular}{ |c|c|c|c| } 
\hline
Transf. type & dCor & RV & GMCC \\
\hline
line Vs circle    & 0.5258 & 0.27378 & 0.871 \\ 
\hline
\end{tabular}
\caption{Similarity distance measurements for line Vs circle}
\label{linvscir}
\end{table}

The distance measured by GMCC is 0.87 while the distance measured by dCor is 0.57. dCor is less sensible to nonlinear transformations than GMCC. our correlation coefficient is a worse tool for measuring nonlinear correlation but at the same time is a better metric for measuring only linear correlations.

\subsection{Clustering robot trajectories}\label{toy_traj}

Six different type of motions were recorded in a KUKA LWR arm. The end effector's cartesian position was recorded for $x$, $y$ and $z$. For each motion type five different demonstrations were recorded with different scales, different starting points and different orientations. The selected motions are lines, U-shape, S-shape, circular,triangle and hello trajectories. All the demonstrations were fixed to the same time length and same amount of time samples. Then, the similarity distance matrix was computed among the demonstrations and this distance matrix was used as the input for a hierarchical clustering.

The similarity matrix of each 30 demonstrations~(~six gestures by five demonstrations) was computed for the three correlation distances dCor, RV and GMCC. The obtained similarity matrices are presented in the Fig.~\ref{fig:siminar_mat}.

The first conclusion is extracted from Fig.~\ref{fig:simmdcor}. dCor distance is measuring small distances not only between the demonstrations of the same motion but also between the demonstrations of different motions. From here, we can predict that at least with the recorded motions, dCor will fail classifying or clustering them correctly. RV coefficient and GMCC are obtaining similar results. For some gestures GMCC differentiates better. For example, between circular trajectories and S-shape trajectories. Anyway, GMCC is getting worse results discriminating linear movements and S-shape movements. It is also relevant to remark how badly the hello trajectories are correlated. None RV coefficient or GMCC is giving good similarity distance between most of the hello trajectories.

\begin{figure*}[t]
\centering
\begin{subfigure}{.33\textwidth}
  \includegraphics[width=\linewidth]{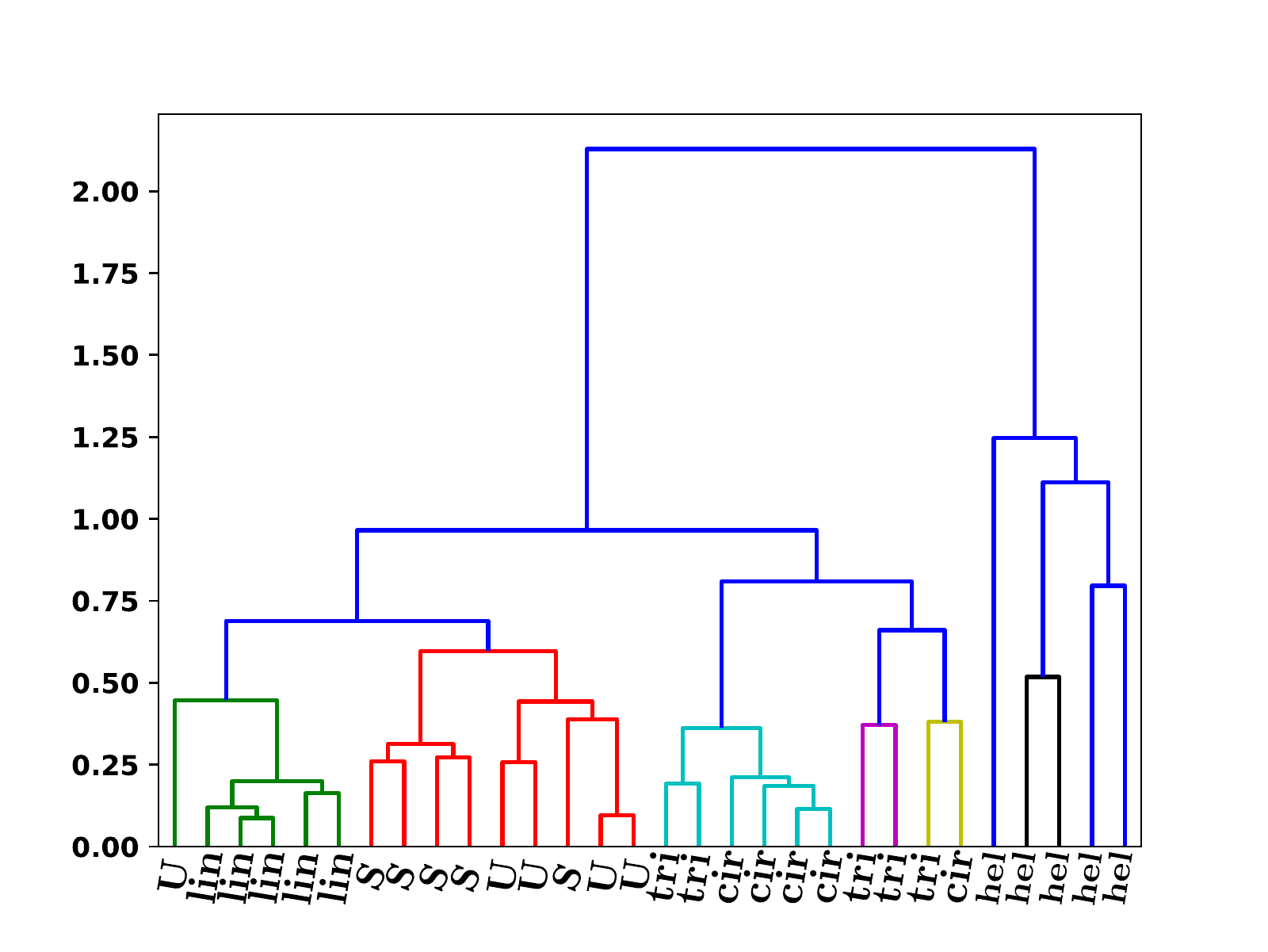}
  \caption{RV's dendrogram}
  \label{fig:dmrv}
\end{subfigure}%
\begin{subfigure}{.33\textwidth}
  \includegraphics[width=\linewidth]{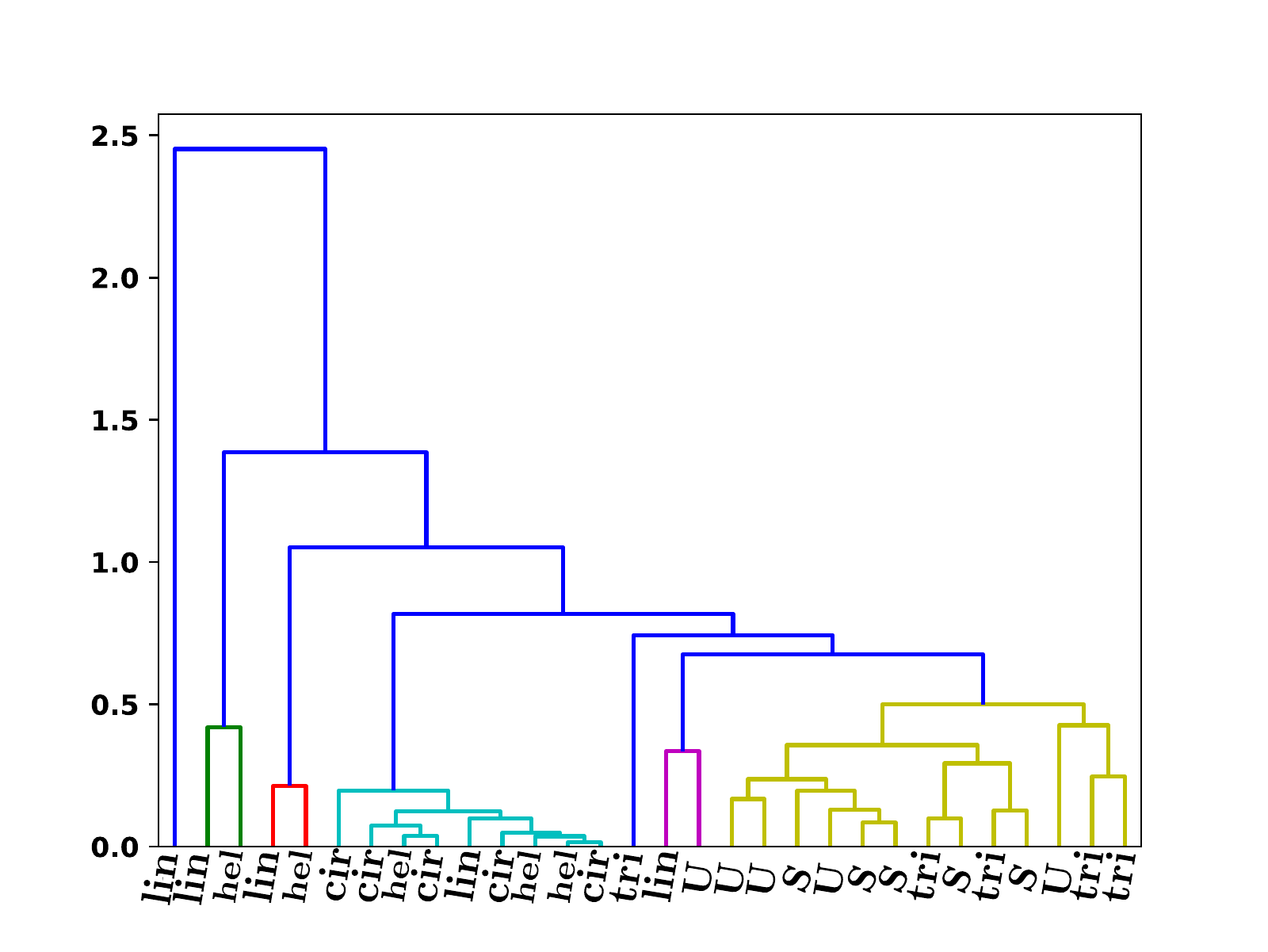}
  \caption{dCor's dendrogram}
  \label{fig:dmdcor}
\end{subfigure}
\begin{subfigure}{.33\textwidth}
  \includegraphics[width=\linewidth]{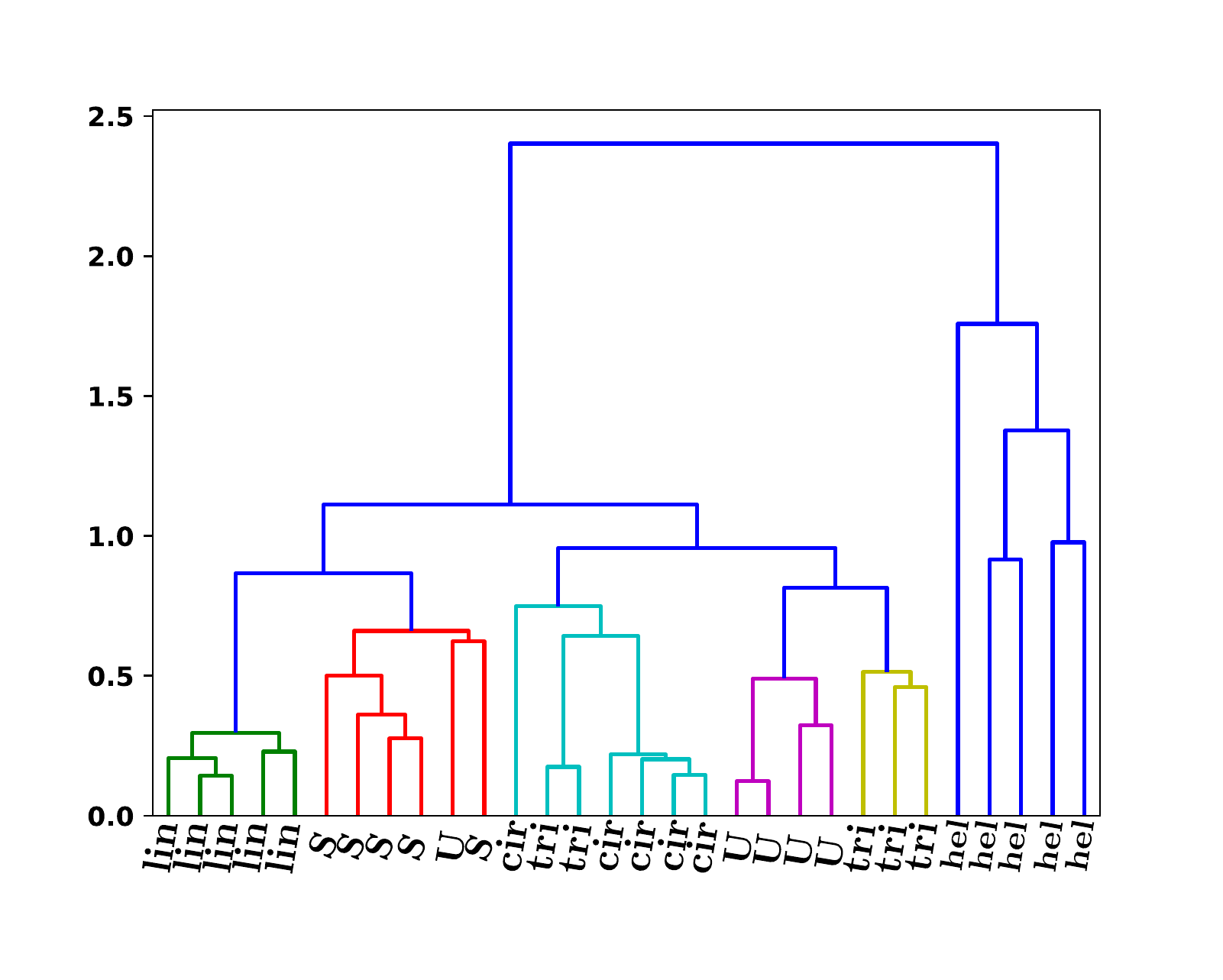}
  \caption{GMCC's dendrogram}
  \label{fig:dmgmcc}
\end{subfigure}%
\caption{Dendrograms computed for RV, dCor and GMCC}
\label{fig:dendom}
\end{figure*}

The recorded demonstrations were clustered by hierarchical clustering to observe the application of our algorithm in a real case scenario. During first evaluation, we considered to cluster the demonstrations in six clusters. However, the hello trajectories were very dissimilar not only in comparison with other trajectories but also between each other as it can be observed in the dendrograms from Fig.~\ref{fig:dendom}. This was making the clustering to fail as the hello trajectories were not clustered in the same cluster, but each one was a unique cluster and then in a unique cluster we could find all the other gestures. In order to solve this, the hierarchical clustering tries to cluster the no-hello motions in five different clusters. 

The applied linkage criterion was the average distance between the clusters.The average linkage criterion will use the average distance between all the elements in both cluster to decide if they should merge or not
$$d(u,v) = \sum_{i=1}^n\sum_{j=1}^m\frac{d(u[i],v[j])}{nm}.$$
From the obtained dendrograms several conclusions can be extracted. Expected from the similarity matrix in Fig.~\ref{fig:simmdcor}, dCor solution is not useful for clustering the recorded human gestures. In Fig.~\ref{fig:dmdcor}, it can be observed how the clusters has been built with dCor. There are two main clusters. The one in yellow, clustering in same cluster some triangle trajectories, U-shape trajectories and S-shape trajectories. The other big clustering some line trajectories, hello trajectories and circular trajectories. This poor results can be understood as dCor finds not only linear but also non-linear correlations between trajectories and so, it will find a small similarity distance between most of the trajectories as they are nonlinearly correlated.

The clustering obtained in GMCC's dendrogram is better than the clustering obtained in RV's dendrogram. There are less missclustered elements in GMCC than in RV. GMCC missclusters two triangle trajectories with the circular trajectories and one U-shape trajectory with the S-shape trajectories. Meanwhile, RV dendrogram is also creating a unique cluster between all U-shape and S-shape trajectories. Also, one of the found clusters by RV is a cluster with only two elements containing one triangle trajectory and one circular trajectory. We can conclude that the best tool for clustering three-dimensional motions is the GMCC.

\section{Discussion and Future Work}\label{conclusions}
We presented a multidimensional correlation coefficient that remains invariant for any linear transformation and only linear transformations. This novel multidimensional has been proved to be a multidimensional extension of the Multiple Correlation Coefficient. The validity of GMCC has been proved in a real robot learning scenario and have showed better clustering capacities than state of art multivariate correlation coefficients such as RV and dCor.

While our similarity distance is invariant to any linear transformation, it cannot deal with time morphing and time shifting. A promising improvement of our algorithm will come from the mixture between DTW and GMCC. Another interesting improvement comes from extending GMCC to probabilistic motions. PROMPs are a promising representation of motions in Imitation Learning and we expect to merge PROMPs with GMCC.

\addtolength{\textheight}{-12cm}   




\bibliographystyle{ieeetr}
\bibliography{bibliography}

\end{document}